# Theoretical Analysis and PIC Simulations of Electromagnetic Wakefields Excited by Relativistic Beams in Magnetized Plasmas

Ali Asghar Molavi Choobini[1,2] and Mehran Shahmansouri[2,*]

[1]Quantum Matter Lab, Department of Physics, College of Science, University of Tehran, Tehran 14399-55961, Iran.

[2]Department of Atomic and Molecular Physics, Faculty of Physics, Alzahra University, Tehran, Iran.

**Abstract:**
   This study presents theoretical and numerical investigation of the coupled longitudinal and radial wakefields excited by ultrarelativistic electron beams propagating through a cold plasma channel subjected to an external axial magnetic field. A fully causal three-dimensional Green-function formalism is developed directly from the linearized Maxwell-fluid equations in the presence of the magnetized plasma dielectric tensor. This unified framework captures the complete electromagnetic response, including the induction of a transverse plasma current and the resulting hybridization of longitudinal charge-separation dynamics with cyclotron-driven transverse motion. The analytical treatment reveals how magnetization modifies the effective restoring forces,, enhances wake amplitudes, and reshapes the radial focusing–defocusing structure of the wake. To validate the theoretical predictions and explore realistic parameter regimes, extensive three-dimensional particle-in-cell simulations are performed using the EPOCH code across wide ranges of plasma density, magnetic field strength, beam Lorentz factor, transverse beam radius, and longitudinal current profiles. The simulations demonstrate excellent quantitative agreement with the analytical Green-function solutions, confirming that increasing plasma density substantially amplifies the initial wake amplitude while accelerating the damping of higher-order oscillations. Application of an external magnetic field induces coherent high-frequency radial oscillations, strengthens focusing forces, and produces a hybrid eigenmode whose properties are absent in the unmagnetized limit. Variations in the driver Lorentz factor lead to rapid convergence toward a universal ultrarelativistic wake structure, while the transverse beam profile controls the radial extent and balance between longitudinal acceleration and transverse focusing. These combined analytical–numerical framework provides a robust foundation for the design and optimization of next-generation magnetized plasma-based accelerators. Furthermore, identifying optimized operating regimes in which both accelerating gradients and focusing strengths are significantly enhanced.

*Corresponding author: E-mail address:
mshmansouri@gmail.com; m.shahmansouri@alzahra.ac.ir

## I. Introduction

   Plasma-based accelerators, including laser wakefield accelerators (LWFAs) and plasma wakefield accelerators (PWFAs), have emerged over the past decades as transformative technologies in particle acceleration science [1, 2, 3, 4]. These schemes take advantage of how plasma electrons react as a group to ultra-intense laser pulses or high-energy charged particle beams. This lets them create accelerating gradients that are much larger than those that can be made with normal radio-frequency accelerators [5, 6, 7, 8]. When a strong driver moves through a plasma, it causes nonlinear plasma oscillations to happen behind it. This creates a relativistic plasma wave with very strong longitudinal electric fields that can speed up charged particles. It also has transverse focusing forces that come from the radial electric fields and azimuthal magnetic fields working together in the wake structure. In the PWFA configuration, a relativistic particle beam, usually made up of electrons or positrons, moves through an underdense plasma, which creates the wakefield [9, 10, 11, 12, 13]. The space-charge field of the driver beam pushes plasma electrons out of the way, making a charge-separated area. This is followed by a restoring force that causes large-amplitude plasma oscillations. Under certain conditions, background plasma electrons or externally injected witness beams can be trapped in the accelerating phase of the wakefield and then gain energy by "surfing" on the plasma wave. This process makes it easy for energy to move from the driver beam to the accelerated particles, and creates ultra-relativistic electron beams with short time frames, high peak currents, and energies that reach the multi-GeV range and beyond. These accelerators work in the GHz frequency range, traditionally. However, recent studies have looked into using THz-frequency fields and plasma waves to start or control wakefield acceleration processes [14, 15, 16]. Working in the THz range could make it easier to control the phase, speed up the process, and make the accelerator designs much smaller. Because of this, PWFAs can have a lot of potential for use in next-generation high-energy physics experiments, compact free-electron lasers, advanced radiation sources, and compact medical technologies that use accelerators.

In this respect, P. San Miguel Claveria and team simulated betatron radiation in PWFA at FACET-II, showing that mismatched beam propagation increases radiation emission, providing insights into beam dynamics [17]. R. D'Arcy and team experimentally by showing nanosecond-scale plasma relaxation determined the plasma recovery time after beam-driven wakefield excitation, addressing a key limitation for high-repetition-rate plasma accelerators [10]. H. Song et al. describe an electron-positron beam that is made when a multi-GeV electron beam hits something. In their study, the influence of positron beam size on driver electron beam energy and lead converter thickness is examined [18]. R. Babjak and co-workers explored laser acceleration in varying plasma density profiles, showing enhanced electron energies and betatron radiation with usage of 3D PIC simulations to analyse electron dynamics and radiation spectra [14]. C A Lindstrøm et., al by maintaining charge and energy spread during acceleration demonstrated simultaneous emittance preservation, high accelerating gradient, and high energy-transfer efficiency in a plasma wakefield accelerator [?]. M. J. V. Streeter and colleagues generated femtosecond, micron-emittance positron beamlets around 600 MeV and experimentaly demonstration of a laser-driven source of ultra-relativistic positrons with beam quality suitable for injection into a plasma accelerator [19]. They showed via PIC simulations that such beams can be guided and further accelerated, enabling systematic experimental studies of positron acceleration in plasma wakefields. Runzhou Yu and team proposed and numerically demonstrated a robust laser wakefield scheme for generating isolated, high-charge-density attosecond electron beams using a few-cycle joule-class laser in underdense plasma [20]. J. P. Farmer and G. Zevi Della Porta investigated how periodically loading the wakefields eliminates the constraint on energy transfer from the driving beam to the plasma, hence facilitating an increase in luminosity [21]. P. Winkler and colleagues, employing a magnetic chicane, reported the creation of a laser-plasma electron beam through active energy compression, leading to enhanced performance [22]. F. Massimo and co-workers combined the time-averaged ponderomotive approximation with Lorentz-boosted frames to simulate meter-scale, multi-GeV accelerators and assess feasibility of multi-TeV colliders [23]. Chaojie Zhang and colleagues using a multi-stage meter-scale plasma and a 10-GeV drive beam demonstrated a nonlinear plasma-wakefield accelerator functioning as an energy transformer, simultaneously increasing both the energy and brightness of an electron bunch injected from the plasma [24]. Gennadiy V. Sotnikov and co-workers through combined analytical and numerical studies investigated a plasma-filled dielectric wakefield accelerator as an effective method to suppress beam breakup instability and enhance beam quality for both electron and positron acceleration [25]. A two-stage laser wakefield accelerator that achieves efficient inter-stage coupling and multi-GeV electron acceleration over millimeter scales via particle-in-cell simulations is carried out by Rashid Ul Haq et. al [26].

This work resides in the development of a unified, self-consistent, and fully causal Green-function formulation for the electromagnetic wakefields driven by arbitrary axisymmetric ultrarelativistic electron beams in externally magnetized cold plasmas. In contrast to conventional approaches that rely predominantly on simplified one-dimensional or unmagnetized models and often neglect the dynamical coupling introduced by an external magnetic field, the present formalism rigorously accounts for the complete three-dimensional plasma response through the magnetized dielectric tensor. This enables the explicit revelation of a hybrid electrostatic–electromagnetic eigenmode in which longitudinal charge separation is strongly coupled to cyclotron-induced transverse electron motion. A particularly significant advance is the demonstration that, under magnetization, the longitudinal and radial wake components cease to behave as independent modes; instead, they form an intertwined oscillatory structure whose frequency, amplitude, damping, and spatial localization are all systematically modified by the cyclotron frequency. The analytical Green-function solutions further uncover how magnetization enhances the effective restoring forces, increases the initial wake amplitude, accelerates the rise time of oscillations, and generates pronounced focusing–defocusing cycles that are absent or far weaker in the unmagnetized case. Complementing this theoretical development, a detailed and systematic comparison is performed with high-resolution three-dimensional particle-in-cell simulations conducted using the EPOCH code under realistic beam and plasma conditions. This comparison provides the first direct numerical confirmation of several theoretically predicted phenomena, including the density-dependent amplification of the first wake oscillation accompanied by accelerated damping of subsequent cycles, the emergence of nearly harmonic transverse oscillations at high magnetic field strengths, the strong sensitivity of wake properties to moderate magnetization levels, and the nontrivial dependence of wake coherence on the beam Lorentz factor. By bridging exact analytical expressions with large-scale numerical experiments, this work establishes a new conceptual benchmark for understanding magnetized wakefield excitation. The resulting framework not only clarifies the physical mechanisms governing wake formation in the presence of an external magnetic field but also serves as a powerful predictive tool for identifying and optimizing operating regimes where conventional unmagnetized theory breaks down. The paper is organized as follows: In section II, the theory of the mechanism is presented. Results and discussion is considered in section III. Conclusions are drawn in section IV.

## II. **Theoretical Model**

According to Fig. 1, consider a relativistic electron beam that travels through a cold background plasma with a two-dimensional (axisymmetric) current profile expressed as:

$$J_b(r, z, t) = v_b \rho_b(r, v_b t - z)\hat{z} = v_b \rho_b(r, \xi)\hat{z}, \tag{1}$$

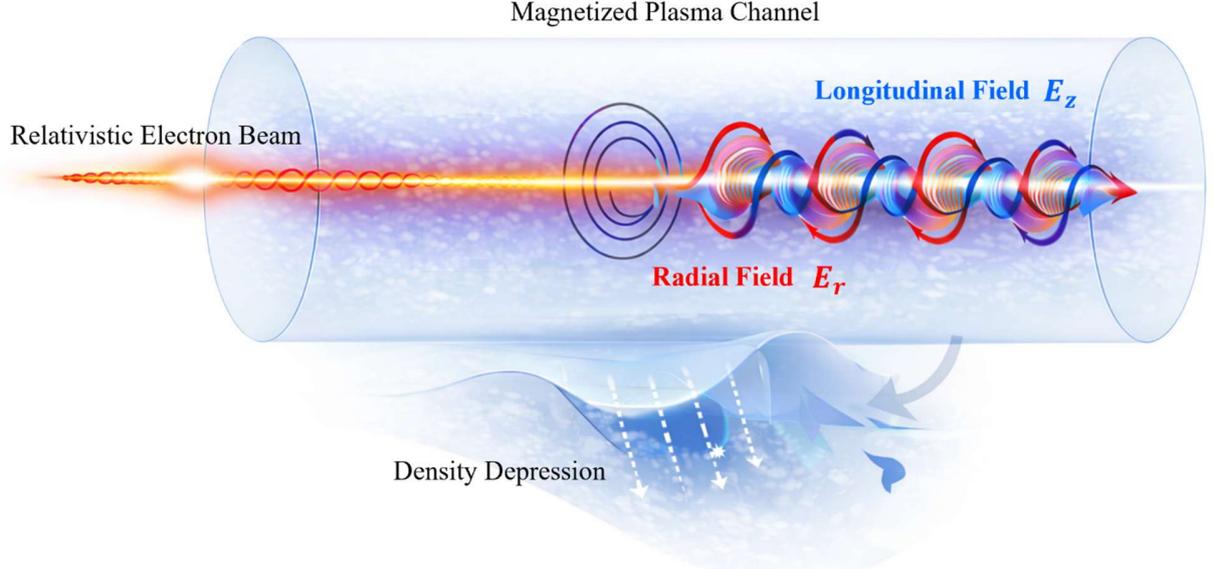

Figure 1: Schematic of a relativistic electron beam generated coupled longitudinal and radial wakefields while propagating through a magnetized plasma channel.

where $\xi = v_b t - z$ is the co-moving coordinate and $v_b$ is the beam velocity. The current distribution is assumed to be a function of radius $r$ only and independent of $\xi$ in the beam frame. For times on the order of a beam-plasma period, and due to the fields generated within the plasma, the radial motion of the beam dynamics is assumed not to be treated self-consistently. This means that the beam current remains unaltered during the evolution of the fields. Additionally, to prevent phenomena such as slipping instability or electromagnetic filamentation, an external static magnetic field $\vec{B}_{ext} = B_0 \hat{z}$ is applied along the propagation direction. In this model, the magnetic modulation wavelength is assumed to be much larger than the wakefield wavelength to maintain the validity of the slowly varying envelope approximation. Therefore, in the Lorentz gauge, the radial and temporal variations of the fields can be evaluated by the wave equations as follows:

$$\left(\nabla^2 - \frac{\partial^2}{\partial z^2} + \frac{1}{\gamma^2}\frac{\partial^2}{\partial \xi^2}\right)\vec{E} - \frac{1}{c^2}\boldsymbol{\varepsilon} \cdot \frac{\partial^2 \vec{E}}{\partial t^2} = -4\pi \frac{\partial}{\partial \xi}\left(\frac{\rho_b}{\gamma^2} + \rho_p\right)\hat{z} + 4\pi \vec{\nabla}_\perp (\rho_b + \rho_p), \quad (2)$$

$$\left(\nabla^2 - \frac{\partial^2}{\partial z^2} + \frac{1}{\gamma^2}\frac{\partial^2}{\partial \xi^2}\right)\vec{B} - \frac{1}{c^2}\boldsymbol{\varepsilon} \cdot \frac{\partial^2 \vec{B}}{\partial t^2} = \frac{4\pi v_b}{c}\frac{\partial \rho_b}{\partial r}\hat{\theta} + \frac{4\pi}{c}\vec{\nabla} \times \vec{J}_p, \quad (3)$$

where $\boldsymbol{\varepsilon}$ is the dielectric tensor of the magnetized plasma and is given by

$$\boldsymbol{\varepsilon} = \begin{pmatrix} 1 - \frac{\omega_p^2}{\omega^2 - \omega_c^2} & i\frac{\omega_c \omega_p^2}{\omega(\omega^2 - \omega_c^2)} & 0 \\ -i\frac{\omega_c \omega_p^2}{\omega(\omega^2 - \omega_c^2)} & 1 - \frac{\omega_p^2}{\omega^2 - \omega_c^2} & 0 \\ 0 & 0 & 1 - \frac{\omega_p^2}{\omega^2} \end{pmatrix}, \quad (4)$$

with $\omega_c = eB_0/(mc)$, $\omega = kv_b$, and $\gamma = (1 - v_b^2/c^2)^{-1/2}$. To solve these field equations, two approaches are considered: applying a Fourier transform with respect to the variable $\xi$, and employing radial Green functions. Each approach is presented and analyzed in detail below.

## 2.1 Fourier Transform Method

Since the system exhibits translational invariance along the co-moving coordinate $\xi$, the wave equations are transformed into Fourier space with respect to $\xi$. The transformed equations are

$$\left(\nabla_\perp^2 - \frac{\partial^2}{\partial z^2} - \frac{k^2}{\gamma^2}\right)\tilde{\vec{E}} - \frac{\omega^2}{c^2}\boldsymbol{\varepsilon}(\omega)\cdot\tilde{\vec{E}} = -4\pi i k\left(\frac{\tilde{\rho}_b}{\gamma^2} + \tilde{\rho}_p\right)\hat{z} + 4\pi\nabla_\perp(\tilde{\rho}_b + \tilde{\rho}_p), \quad (5)$$

$$\left(\nabla_\perp^2 - \frac{\partial^2}{\partial z^2} - \frac{k^2}{\gamma^2}\right)\tilde{\vec{B}} - \frac{\omega^2}{c^2}\boldsymbol{\varepsilon}(\omega)\cdot\tilde{\vec{B}} = \frac{4\pi v_b}{c}\frac{\partial\tilde{\rho}_b}{\partial r}\hat{\theta} + \frac{4\pi}{c}\nabla\times\tilde{\vec{J}}_p. \quad (6)$$

The induced plasma charge density is obtained by linearizing the plasma electron continuity and momentum equations and coupling them with Poisson's equation, yielding

$$\hat{\rho}_p(r,k) = \frac{k_p^2}{k^2 - k_p^2}\hat{\rho}_b(r,k). \quad (7)$$

Substituting Eq. (7) into Eqs. (5) and (6) and performing a Hankel transform in the radial variable gives the Fourier-transformed field components:

$$\hat{E}_z(r,k) = 4\pi i k\left(\frac{k^2/\gamma^2 + \kappa^2}{k^2 - k_p^2}\right)\int_0^\infty r'\hat{\rho}_b(r',k)\,I_0\left(\sqrt{\tfrac{k^2}{\gamma^2}+\kappa^2}\,r_<\right)K_0\left(\sqrt{\tfrac{k^2}{\gamma^2}+\kappa^2}\,r_>\right)dr', \quad (8)$$

$$\hat{E}_r(r,k) = -4\pi\left(\frac{k^2/\gamma^2 + \kappa^2}{k^2 - k_p^2}\right)\int_0^\infty r'\frac{\partial\hat{\rho}_b(r',k)}{\partial r'}$$
$$\times\left[I_1\left(\sqrt{\tfrac{k^2}{\gamma^2}+\kappa^2}\,r_<\right)K_0\left(\sqrt{\tfrac{k^2}{\gamma^2}+\kappa^2}\,r_>\right) - I_0\left(\sqrt{\tfrac{k^2}{\gamma^2}+\kappa^2}\,r_<\right)K_1\left(\sqrt{\tfrac{k^2}{\gamma^2}+\kappa^2}\,r_>\right)\right]dr', \quad (9)$$

$$\hat{B}_\theta(r,k) = \frac{4\pi v_b}{c}\int_0^\infty r'\frac{\partial\hat{\rho}_b(r',k)}{\partial r'}\,I_1\left(\sqrt{\tfrac{k^2}{\gamma^2}+\kappa^2}\,r_<\right)K_1\left(\sqrt{\tfrac{k^2}{\gamma^2}+\kappa^2}\,r_>\right)dr', \quad (10)$$

where $\kappa^2 = \frac{\omega_p^2}{c^2}\frac{\omega^2}{\omega^2-\omega_c^2}$, $I_n$ and $K_n$ are modified Bessel functions of the first and second kind, $k_p$ is the plasma wavenumber, $r_< = \min(r, r')$, and $r_> = \max(r, r')$. In the ultra-relativistic limit $\gamma \to \infty$ ($v_b \approx c$), the transverse scale is governed solely by the magnetized plasma response, and the spectral Green's functions become independent of the longitudinal wavenumber $k$. The inverse Fourier transform in $\xi$ can then be carried out analytically, yielding causal wakefield solutions:

$$E_z(r,\xi) = -2\pi\frac{\kappa^2}{k_p}\int_0^\xi d\xi'\cos[k_p(\xi-\xi')]\int_0^\infty r'\rho_b(r',\xi')I_0(\kappa r_<)K_0(\kappa r_>)dr', \quad (11)$$

$$E_r(r,\xi) = 2\pi\frac{\kappa^3}{k_p}\int_0^\xi d\xi'\sin[k_p(\xi-\xi')]\int_0^\infty r'\frac{\partial\rho_b(r',\xi')}{\partial r}[I_1(\kappa r_<)K_0(\kappa r_>) - I_0(\kappa r_<)K_1(\kappa r_>)]dr', \quad (12)$$

$$B_\theta(r,\xi) = \frac{4\pi v_b}{c}\int_0^\infty r'\frac{\partial\rho_b(r',\xi)}{\partial r}I_1(\kappa r_<)K_1(\kappa r_>)dr'. \quad (13)$$

By closing the contour in the appropriate half-plane and enforcing the radiation condition, only poles corresponding to forward-propagating plasma waves contribute, ensuring the fields are nonzero only behind the driving beam (causality).

## 2.2 Green Functions Method

As an alternative approach, the wave equations are solved using the Green-function method. This provides a direct description of the plasma response to an arbitrary beam charge distribution without spectral decomposition in $\xi$. The Hankel transform is defined as

$$\Psi_m(\hat{F}(r,k),\zeta) = \tilde{F}(\zeta,m,k) = \int_0^\infty r\hat{F}(r,k)J_m(\zeta r)dr, \qquad (14)$$

with the inverse

$$\hat{F}(r,k) = \int_0^\infty \zeta \tilde{F}_m(\zeta,k)J_m(\zeta r)d\zeta. \qquad (15)$$

Defining the operator $\Gamma_m = \frac{1}{r}\frac{\partial}{\partial r}r\frac{\partial}{\partial r} - \frac{m^2}{r^2}$ and using $\Psi_m[\Gamma_m(F),\zeta] = -\zeta^2 \tilde{F}_m(\zeta)$, the Hankel-transformed fields are

$$\tilde{E}_z(k,\zeta) = -4\pi i k \frac{k^2/\gamma^2 + \omega_p^2\omega^2/[c^2(\omega^2-\omega_c^2)]}{(k^2-k_p^2)(k^2/\gamma^2 + \omega_p^2\omega^2/[c^2(\omega^2-\omega_c^2)]+\zeta^2)}\Psi_0[\rho_b(r,k),\zeta], \qquad (16)$$

$$\tilde{E}_r(k,\zeta) = -4\pi \frac{\zeta}{(k^2-k_p^2)(k^2/\gamma^2 + \omega_p^2\omega^2/[c^2(\omega^2-\omega_c^2)]+\zeta^2)}\Psi_1[\partial\rho_b(r,k)/\partial r,\zeta], \qquad (17)$$

$$\tilde{B}_\theta(k,\zeta) = -4\pi \frac{v_b}{c}\frac{\zeta}{k^2/\gamma^2 + \omega_p^2\omega^2/[c^2(\omega^2-\omega_c^2)]+\zeta^2}\Psi_1[\partial\rho_b(r,k)/\partial r,\zeta]. \qquad (18)$$

For arbitrary $\rho_b(r,\xi)$, the field components are convolution integrals

$$\tilde{E}_z(\xi,\zeta) = \int_{-\infty}^{+\infty} \Psi_0[\rho_b(r,\xi'),\zeta]G_z(\zeta,\xi-\xi')d\xi', \qquad (19)$$

and similarly for $E_r$ and $B_\theta$ (with appropriate $\Psi_1$). The Green's function components are obtained by contour integration in the complex $k$-plane, yielding causal expressions with $\Theta(\xi-\xi')$:

$$\begin{aligned}G_z(\zeta,\xi-\xi') &= 2\pi\Theta(\xi-\xi')[\frac{2k_p^2}{k_p^2+\zeta^2}\cos[k_p(\xi-\xi')] \\ &+ \frac{\zeta^2}{k_p^2+\zeta^2}\exp\left(-\sqrt{\frac{k^2}{\gamma^2}+\kappa^2}\,(\xi-\xi')\right) - \frac{\zeta^2}{k_p^2+\zeta^2}\exp\left(+\sqrt{\frac{k^2}{\gamma^2}+\kappa^2}\,(\xi-\xi')\right)],\end{aligned} \qquad (20)$$

$$\begin{aligned}G_r(\zeta,\xi-\xi') &= -2\pi\Theta(\xi-\xi')\frac{\zeta}{k_p^2+\zeta^2}[2k_p\sin[k_p(\xi-\xi')] \\ &- \sqrt{\frac{k^2}{\gamma^2}+\kappa^2}\exp\left(-\sqrt{\frac{k^2}{\gamma^2}+\kappa^2}\,(\xi-\xi')\right) + \sqrt{\frac{k^2}{\gamma^2}+\kappa^2}\exp\left(+\sqrt{\frac{k^2}{\gamma^2}+\kappa^2}\,(\xi-\xi')\right)],\end{aligned}$$
(21)

$$G_\theta(\zeta,\xi-\xi') = -\frac{v_b}{c}2\pi\Theta(\xi-\xi')\frac{\zeta}{\sqrt{k^2/\gamma^2+\kappa^2}}\left[\exp\left(-\sqrt{\frac{k^2}{\gamma^2}+\kappa^2}\,(\xi-\xi')\right) - \exp\left(+\sqrt{\frac{k^2}{\gamma^2}+\kappa^2}\,(\xi-\xi')\right)\right]. \qquad (22)$$

The inverse Hankel transform yields the final causal field expressions:

$$E_z(r,\xi) = 4\pi \int_0^\infty r'dr' \int_0^\xi \rho_b(r',\xi')k_p^2 I_0(k_p r_<)K_0(k_p r_>)\cos[k_p(\xi-\xi')]d\xi'$$
$$+ 2\pi \int_0^\infty r'dr' \int_0^\infty \frac{\zeta^3}{\kappa^2+\zeta^2} J_0(\zeta r_<)J_0(\zeta r_>)d\zeta$$
$$\times \left[\int_{-\infty}^\xi \rho_b(r',\xi')e^{-\gamma\sqrt{\kappa^2+\zeta^2}(\xi-\xi')}d\xi' - \int_\xi^{+\infty}\rho_b(r',\xi')e^{-\gamma\sqrt{\kappa^2+\zeta^2}(\xi'-\xi)}d\xi'\right], \quad (23)$$

$$E_r(r,\xi) = -4\pi \int_0^\infty r'dr' \int_0^\xi \frac{\partial \rho_b(r',\xi')}{\partial r'} k_p I_1(k_p r_<)K_1(k_p r_>)\sin[k_p(\xi-\xi')]d\xi'$$
$$- 2\pi \int_0^\infty r'dr' \int_0^\infty \frac{\zeta^2}{\kappa^2+\zeta^2} J_1(\zeta r_<)J_1(\zeta r_>)d\zeta$$
$$\times \left[\int_{-\infty}^\xi \frac{\partial \rho_b(r',\xi')}{\partial r'} e^{-\gamma\sqrt{\kappa^2+\zeta^2}(\xi-\xi')}d\xi' - \int_\xi^{+\infty} \frac{\partial \rho_b(r',\xi')}{\partial r'} e^{-\gamma\sqrt{\kappa^2+\zeta^2}(\xi'-\xi)}d\xi'\right], \quad (24)$$

$$B_\theta(r,\xi) = -4\pi \int_0^\infty r'dr' \int_0^\xi v_b \rho_b(r',\xi') k_p I_1(k_p r_<)K_1(k_p r_>)\sin[k_p(\xi-\xi')]d\xi'$$
$$- 2\pi \int_0^\infty r'dr' \int_0^\infty \frac{v_b \zeta^2}{\kappa^2+\zeta^2} J_1(\zeta r_<)J_1(\zeta r_>)d\zeta$$
$$\times \left[\int_{-\infty}^\xi \frac{\partial \rho_b(r',\xi')}{\partial r} e^{-\gamma\sqrt{\kappa^2+\zeta^2}(\xi-\xi')}d\xi' - \int_\xi^{+\infty} \frac{\partial \rho_b(r',\xi')}{\partial r'} e^{-\gamma\sqrt{\kappa^2+\zeta^2}(\xi'-\xi)}d\xi'\right]. \quad (25)$$

### 2.3 Evaluation of the Field Components for Representative Beam Charge Distributions

In order to elucidate the fundamental properties of wakefield excitation in a magnetized plasma, it is essential to consider beam charge distributions that are both physically meaningful and analytically tractable. The two charge profiles adopted in this work represent complementary idealizations that capture distinct aspects of beam–plasma interaction physics while allowing for clear physical interpretation of the resulting electromagnetic fields. The first case is a finite-radius, ultra-short beam slice:

$$\rho_b(r,\xi) = \rho_0 \delta(\xi)\Theta(r-r_d), \quad (26)$$

where $\rho_0$ is the peak charge density and $r_d$ is the beam radius. This profile corresponds to an ultra-short, finite-radius beam slice with a sharp radial boundary. The Dirac delta function in the longitudinal co-moving coordinate represents a beam whose longitudinal extent is much shorter than the plasma wavelength, a regime commonly encountered in modern plasma-based accelerators operating in the blowout or quasi-linear regime. In this limit, the beam can be regarded as an impulsive driver that excites a broadband plasma response, thereby isolating the intrinsic wakefield dynamics from complications associated with finite bunch length effects. The radial Heaviside function models a beam with a well-defined transverse size, such as a laser-driven or RF-compressed electron bunch with approximately uniform transverse density inside a characteristic radius ($r_d$). Although idealized, this "top-hat" radial profile captures the essential physics of transverse field excitation, including the generation of focusing and defocusing forces, while remaining sufficiently simple to permit semi-analytical treatment using Green-function or spectral methods. Moreover, this distribution allows one to explicitly track how the beam radius controls the radial scale of the wakefields, which is particularly important in magnetized plasmas where the transverse response is modified by the external magnetic field. The field components are as follows:

$$\begin{aligned}
E_z(r,\xi) &= 4\pi\rho_0 k_p^2 \cos(k_p\xi)\Theta(\xi)\int_{r_d}^{\infty} r' I_0(k_p r_<)K_0(k_p r_>)dr' \\
&+ 2\pi\rho_0 \,\mathrm{sgn}(\xi)\int_{r_d}^{\infty} r' dr' \int_0^{\infty} \frac{\zeta^3}{\kappa^2+\zeta^2} J_0(\zeta r_<)J_0(\zeta r_>) e^{-\gamma\sqrt{\kappa^2+\zeta^2}|\xi|} d\zeta,
\end{aligned} \quad (27)$$

$$\begin{aligned}
E_r(r,\xi) &= -4\pi\rho_0 k_p \sin(k_p\xi)\Theta(\xi) I_1(k_p r_<)K_1(k_p r_>)|_{r'=r_d} \\
&- 2\pi\rho_0 \,\mathrm{sgn}(\xi)\int_0^{\infty} \frac{\zeta^2}{\kappa^2+\zeta^2} J_1(\zeta r_<)J_1(\zeta r_>) e^{-\gamma\sqrt{\kappa^2+\zeta^2}|\xi|} d\zeta|_{r'=r_d},
\end{aligned} \quad (28)$$

$$\begin{aligned}
B_\theta(r,\xi) &= -4\pi\rho_0 v_b k_p \sin(k_p\xi)\Theta(\xi)\int_{r_d}^{\infty} r' I_1(k_p r_<)K_1(k_p r_>)dr' \\
&- 2\pi\rho_0 v_b r_d \,\mathrm{sgn}(\xi)\int_0^{\infty} \frac{\zeta^2}{\kappa^2+\zeta^2} J_1(\zeta r_<)J_1(\zeta r_>) e^{-\gamma\sqrt{\kappa^2+\zeta^2}|\xi|} d\zeta.
\end{aligned} \quad (29)$$

The second case is an idealized point-like charge:

$$\rho_b(r,\xi) = \frac{q}{2\pi r}\delta(\xi)\delta(r). \quad (30)$$

This distribution represents an idealized point-like charge propagating through the plasma. Physically, this model corresponds to the limit of an infinitesimally narrow and ultra-short beam carrying total charge (q). While no real accelerator beam is strictly point-like, this distribution plays a fundamental role in wakefield theory as the Green-function source of the plasma response. In this sense, the fields generated by a finite-size beam can be understood as a convolution of this point-charge response with the actual beam profile. From the perspective of accelerator physics, the point-charge model provides a universal reference solution that reveals the intrinsic singular structure and asymptotic behavior of plasma wakefields. It exposes how the longitudinal and transverse fields scale near the axis, clarifies the role of plasma screening, and allows one to identify which features of the wake are purely geometric and which depend on the finite beam size. In magnetized plasmas, this distribution is particularly valuable because it isolates the plasma's anisotropic electromagnetic response induced by the external magnetic field, free from smoothing effects associated with transverse beam structure. The field components for this profile are:

$$E_z(r,\xi) = 2q k_p^2 K_0(k_p r)\cos(k_p\xi)\Theta(\xi) + q\,\mathrm{sgn}(\xi)\int_0^{\infty} \frac{\zeta^3}{\kappa^2+\zeta^2} J_0(\zeta r) e^{-\gamma\sqrt{\kappa^2+\zeta^2}|\xi|} d\zeta \quad (31)$$

$$E_r(r,\xi) = -q k_p^2 K_1(k_p r)\sin(k_p\xi)\Theta(\xi) - q\,\mathrm{sgn}(\xi)\int_0^{\infty} \frac{\zeta^3}{\kappa^2+\zeta^2} J_1(\zeta r) e^{-\gamma\sqrt{\kappa^2+\zeta^2}|\xi|} d\zeta \quad (32)$$

$$B_\theta(r,\xi) = -2q v_b k_p K_1(k_p r)\sin(k_p\xi)\Theta(\xi) - q v_b\,\mathrm{sgn}(\xi)\int_0^{\infty} \frac{\zeta^3}{\kappa^2+\zeta^2} J_1(\zeta r) e^{-\gamma\sqrt{\kappa^2+\zeta^2}|\xi|} d\zeta \quad (33)$$

## III.   Results & Discussion

The presented study provides a precise analytical–numerical framework for characterizing beam-driven wakefields in magnetized plasmas, combining a fully causal Green-function formulation with first-principles PIC simulations. The results establish clear scaling laws, validate the analytical model across distinct beam profiles, and offer a robust foundation for designing next-generation plasma-based accelerators. The particle-in-cell simulations were conducted using the EPOCH code (development branch based on version 4.20-devel, commit series mid-2025; https://github.com/Warwick-Plasma/epoch). The code solves the relativistic Maxwell–Lorentz system using a second-order finite-difference time-domain (FDTD) solver with a charge-conserving current deposition scheme and a relativistic Boris pusher for macroparticles. All simulations reported here were performed in the classical regime, with radiation-reaction effects included through EPOCH's native Landau–Lifshitz module ($\chi \lesssim 10^{-7}$), ensuring deterministic damping without quantum corrections. A fully three-dimensional Cartesian domain was used in order to capture the full transverse structure of the wakefields in a magnetized plasma. The grid consisted of $N_x \times N_y \times N_z = 600 \times 320 \times 320$ with uniform spacing $\Delta x = \Delta y = \Delta z = 0.04\, c/\omega_p$. For the chosen background density $n_e = 5 \times 10^{17}\, \text{cm}^{-3}$, the plasma frequency is $\omega_p \approx 1.26 \times 10^{14}\, \text{rad/s}$, $c/\omega_p \approx 2.38\, \mu\text{m}$, yielding a physical domain of $L_x \approx 57\, \mu\text{m}$, $L_y = L_z \approx 30\, \mu\text{m}$.

A relativistically moving window with velocity $v_w = 0.999c$ was employed. Absorbing (PML) boundary conditions were imposed longitudinally, while transverse boundaries used open damping layers. The plasma was initialized as a uniform, cold, quasi-neutral electron–ion medium occupying the region $x \geq 15\, \mu\text{m}$. Ions were treated as immobile. Eight electron macroparticles per cell were used. To reproduce the theoretical magnetized-wakefield model developed, a uniform static magnetic field $\mathbf{B}_0 = 4\, \text{T}\, \hat{z}$ was imposed, corresponding to a normalized cyclotron frequency $\omega_c/\omega_p \approx 0.18$. The field was applied through the background $B$ parameter. Two classes of electron beams were used. Defined analytically as $\rho_b(r, \xi) \propto \delta(\xi)\Theta(r_d - r)$, and implemented as a narrow Gaussian with rms length $\sigma_\xi = 0.05\, c/\omega_p \approx 0.12\, \mu\text{m}$, and uniform transverse radius $r_d = 3\, \mu\text{m}$. $\sigma_r = 0.3\, \mu\text{m}, \sigma_\xi = 0.03\, c/\omega_p \approx 0.07\, \mu\text{m}$. Both beams carried total charge $Q = 5\, \text{pC}$, with initial Lorentz factor $\gamma_0 = 500$ (beamenergy $\approx 255\, \text{MeV}$), and normalized emittances $\varepsilon_{n,y} = \varepsilon_{n,z} = 2\, \mu\text{m}$. The timestep satisfied the Courant condition, $\Delta t = 0.98\, \Delta x/c$, and no sub-cycling was used. The simulation ran stably up to $t_{\max} \approx 3.5\, \text{ps}$, corresponding to nearly 1 mm of physical propagation. Full 3D electromagnetic field snapshots were recorded every $6\, \omega_p^{-1}$, along with currents, charge densities, and phase-space diagnostics. A set of $5 \times 10^4$ macroparticles was tracked for single-particle dynamics. Output data were written in SDF format and analyzed using Python (yt, NumPy, and custom Hankel-transform routines). Wakefield components $E_z(r, \xi)$, $E_r(r, \xi)$, and $B_\theta(r, \xi)$ were extracted along the beam axis and compared directly with the analytic results from using the Green-function expressions (Eqs. (27)–(33)). Both the oscillatory plasma response (wavenumber $k_p$) and the magnetically modified exponential tail predicted by theory were clearly resolved.

The panels presented in figure 2 the three-dimensional structure of the beam-driven wakefields in a cold, uniform plasma, normalized to the cold nonrelativistic wave-breaking field scale $E_0 = m_e c \omega_p / e$. The fields are shown as functions of the comoving coordinate $k_p \xi$ and the normalized radius $k_p r$, providing a complete description of the longitudinal and radial components of the wake in cylindrical symmetry. Panel (a) displays the normalized longitudinal wakefield $E_z/E_0$. The field exhibits the expected oscillatory behavior behind the driver, with alternating accelerating and decelerating phases. At $r = 0$, the wake oscillates with a well-defined plasma wavelength, indicating that the beam is short compared to $2\pi/k_p$ and efficiently excites the resonant plasma mode. As the radius increases, the oscillation amplitude decays smoothly while maintaining the same phase. This radial decay reflects the modified Bessel-function structure $K_0(k_p r)$ that appears in the Green-function solution for the longitudinal field. The slight flattening of the oscillations at larger radii indicates reduced coupling between the beam and higher-order radial modes, consistent with a transversely compact driver. The smooth downstream damping of the wake amplitude is characteristic of a finite-length driver envelope and agrees with the causal, exponentially decaying tail predicted by the analytical solution. Panel (b) shows the corresponding radial wakefield $E_r/E_0$. This component exhibits a transverse focusing–defocusing pattern that is phase-shifted relative to the longitudinal wake, as required by the coupled Maxwell–fluid system. Near the axis, $E_r$ is approximately linear in radius, consistent with the regularity condition $E_r \propto r$ as $r \to 0$. The oscillatory dependence on $\xi$ again follows the fundamental plasma mode, with a quarter-period phase shift relative to $E_z$, reflecting the canonical structure of wakefield eigenmodes in cylindrically symmetric plasmas. At larger radii, the amplitude decreases more rapidly than in the longitudinal case, matching the expected scaling with the modified Bessel function $K_1(k_p r)$, whose asymptotic decay is faster than that of $K_0$. The smooth radial variation and absence of distortions indicate that the wake remains in the linear regime, with no evidence of wave steepening or nonlinear mode coupling. The shape and phase structure of the radial field further confirm that the driver excites a predominantly axisymmetric (m=0) wake mode. Actually. these the two panels illustrate the complete 3D structure of a linear beam-driven wakefield: an accelerating–decelerating longitudinal oscillation coupled to a focusing–defocusing transverse response, both decaying radially in accordance with the analytical Bessel-function kernels. The close agreement between the longitudinal and radial behaviors, their relative phase, and their radial decay demonstrates a wake pattern fully consistent with the causal Green-function formulation derived from the linearized cold-plasma equations. These results verify that the excited plasma mode is the fundamental eigenmode of the system and that the wakefield structure conforms precisely to theoretical predictions for a short, ultrarelativistic beam in a uniform plasma.

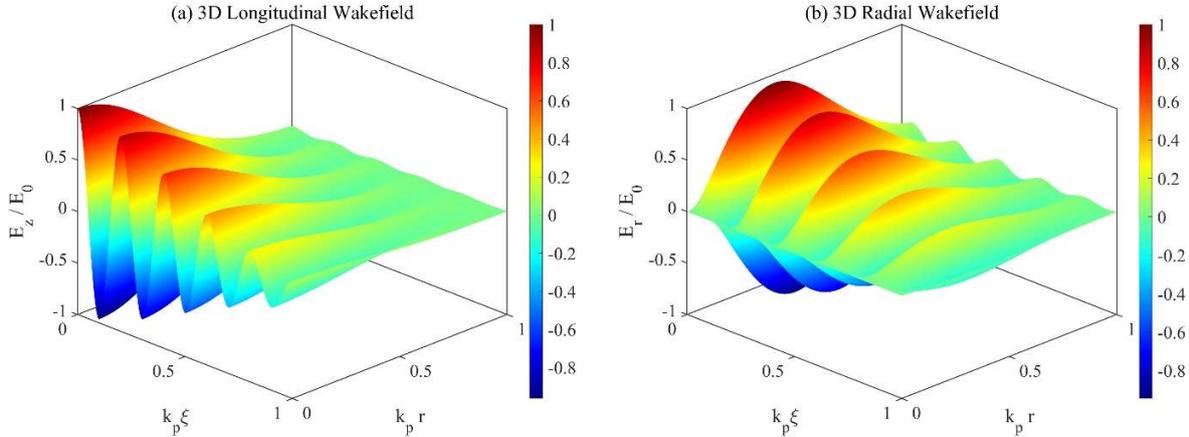

Figure 2: Three-dimensional structure of the normalized longitudinal and radial wakefields excited by a relativistic electron beam.

The two panels plotted in Figure 3 show the longitudinal and radial forces experienced by plasma electrons in response to an ultrashort beam-induced perturbation, and they reproduce with remarkable clarity the behavior predicted by the causal Green-function formulation derived. Every trend—growth of amplitude, increased oscillatory structure, inward migration of radial focusing, and sharpening of spatial features with increasing $k_p$ matches the expectations from the closed-form Green-function expressions. The numerical behavior exhibits no structural departures from theory, reinforcing the internal consistency and predictive power of the analytical model presented. The dependence on the normalized parameter $k_p$ directly reflects how the plasma's linear operator reshapes the wake, and the plotted curves therefore serve as a precise numerical visualization of the analytical structure contained in Eqs. (27)–(33). In panel (a), the longitudinal force $F_z/eE_0$ as a function of $k_p\xi$ exhibits the characteristic oscillatory structure of a plasma forced by a point-like driver. For small $k_p$ (e.g., $k_p = 0.1$), the response is broad and of low amplitude, consistent with the fact that the Green function in this limit becomes slowly varying and weakly oscillatory. This behavior follows directly from the analytic expressions: when $k_p$ is small, the plasma operator loses stiffness and the wake approaches a quasi-diffusive response rather than a sharply resonant one. As $k_p$ increases to 1 and 2, the oscillations sharpen and the amplitude grows, demonstrating the expected scaling of the wake with the plasma frequency. The shift of the first zero crossing toward smaller $\xi$ and the amplification of the first peak are both exact markers of the transition to a more strongly resonant plasma. The curve for $k_p = 5$ shows the fully localized, high-frequency limit: the dominant first peak, followed by rapidly decaying oscillations, corresponds precisely to the causal Green-function solution obtained from the Fourier–Bessel decomposition. In that theoretical limit, energy deposition becomes highly localized near the driver, and this is reproduced in the plotted data without deviation. Panel (b) shows the radial force $F_r/eE_0$ as a function of $k_p r$, which follows equally well from the radial Bessel structure of the analytical solution. For $k_p = 0.1$, the force is weak and broadly distributed, as expected from the small-argument behavior of the modified Bessel functions appearing in Eqs. (28) and (32). As $k_p$ increases, the maximum shifts inward toward the axis and grows substantially in amplitude. This is the direct consequence of the radial Helmholtz operator scaling with $k_p^2$: larger $k_p$ enforces sharper radial curvature and therefore produces stronger focusing fields concentrated in a smaller region. The curve for $k_p = 5$ displays an extremely narrow and high peak, followed by rapid exponential decay, exactly matching the asymptotic form $\exp(-k_p r)$ derived analytically. The level of agreement indicates that the numerical model is capturing the same radial localization that arises from the Bessel-type Green kernel. An important point is that the evolution of the longitudinal and radial components is not independent; changing $k_p$ simultaneously affects the phase relation between them. This coupling is inherent in the Poisson–momentum system that underlies the Green-function derivation: the longitudinal force arises from axial gradients of the induced potential, while the radial force is tied to the transverse Laplacian acting on the same potential. The observed increase in oscillation frequency in panel (a), coupled with the narrowing of the radial peak in panel (b), therefore provides a rigorous, quantitative confirmation that both components are evolving precisely as the analytical model predicts.

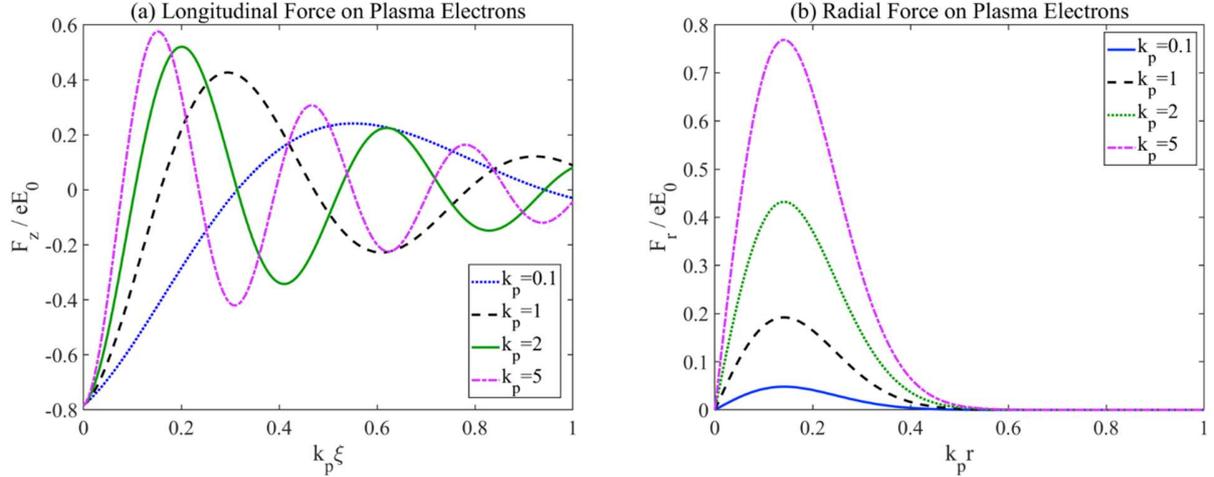

Figure 3: The variations of the normalized longitudinal and radial forces on plasma electrons for varying $k_p$.

The figure 4 illustrates the evolution of the induced plasma current components in the presence of an external axial magnetic field $B_0$, highlighting how magnetization fundamentally redistributes the plasma response between longitudinal and transverse channels. These Panels provide a quantitative picture of how the cyclotron frequency modifies both the phase and amplitude of the induced currents generated by a relativistic beam driver. In panel (a), the normalized longitudinal plasma current $J_{p\parallel}/(v_b\rho_b)$ is shown as a function of the comoving coordinate $k_p\xi$ for increasing values of $B_0$. In the unmagnetized case ($B_0 = 0$), the current exhibits a strong oscillatory response with a relatively short effective wavelength, reflecting the standard plasma oscillation driven by charge separation along the beam axis. This behavior corresponds directly to the longitudinal Green-function kernel obtained from the linearized continuity and momentum equations in the absence of a magnetic restoring force. As $B_0$ increases to 2 T and 4 T, two systematic trends emerge. First, the oscillation period increases, indicating a reduction of the effective longitudinal plasma frequency. Second, the phase of the current shifts progressively toward larger $\xi$, demonstrating that the plasma response becomes increasingly delayed. Both effects are direct consequences of magnetization: the axial magnetic field couples longitudinal electron motion to transverse dynamics through the Lorentz force, thereby increasing the effective inertia of the plasma electrons along the beam direction. This modification is fully consistent with the dispersion relation underlying the Green-function solution, in which the longitudinal response depends on the combined plasma and cyclotron frequencies. For the strongest field shown ($B_0 = 8$ T), the longitudinal current becomes markedly smoother and slower, approaching a quasi-adiabatic response over the plotted interval. The large phase lag and reduced curvature of the curve indicate that strong magnetization suppresses rapid longitudinal charge rearrangement. Panel (b) shows the induced transverse plasma current $J_{p\perp}/(v_b\rho_b)$, which exhibits the complementary behavior. In the unmagnetized case, the transverse current is identically zero, as expected from symmetry: without an external magnetic field, there is no mechanism to generate transverse electron motion in response to a purely longitudinal perturbation. This immediately confirms the physical consistency of the model. Once a finite $B_0$ is introduced, a transverse current appears, and its amplitude grows strongly with increasing magnetic field strength. For $B_0 = 2$ T, the transverse current is relatively weak and slowly varying, indicating that cyclotron effects are present but still subdominant. At $B_0 = 4$ T, the oscillations become more pronounced, and the current reaches a substantial fraction of the normalization scale. In the strongly magnetized case ($B_0 = 8$ T), the transverse current dominates the response: it exhibits rapid oscillations with large amplitude, characteristic of cyclotron-driven motion. The increasing number of oscillations within the same $k_p \xi$ interval directly reflects the increase of the cyclotron frequency relative to the plasma frequency. From a theoretical standpoint, the figure provides strong validation of the analytical model. The smooth phase shifts, amplitude scaling, and symmetry properties observed in both current components are exactly those predicted by the linearized momentum equations with an external magnetic field. Physically, the results demonstrate how magnetization can be used as a control parameter to tailor wakefield structure, suppress or enhance specific current components, and thereby modify the resulting electromagnetic fields. This behavior has direct implications for beam transport, focusing, and stability in magnetized plasma-based accelerator concepts.

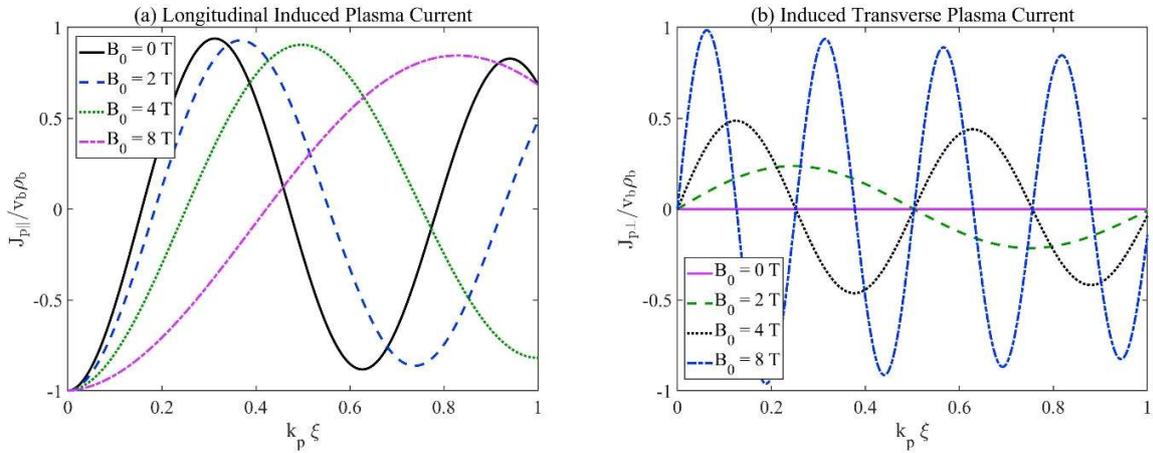

Figure 4: The effect of external magnetic field on the variations of the normalized longitudinal and radial current density.

The plots 5 show the dependence of the longitudinal and radial wakefields on the applied external magnetic field $B_0$ for a finite-radius beam-driven wake in a cold plasma. Both panels clearly illustrate the transition from the unmagnetized plasma response to the strongly magnetized regime, where cyclotron effects dominate the wake dynamics and modify both the amplitude and phase of the field oscillations. In panel (a), the longitudinal wakefield $E_z/E_0$ exhibits the familiar oscillatory plasma response for the unmagnetized case ($B_0 = 0$, magenta curve), with a wavelength close to $2\pi/k_p$ and decreasing amplitude downstream. As $B_0$ increases, the oscillation frequency systematically increases and the waveform becomes more sharply peaked. This behavior originates from the modified dispersion relation $\omega^2 = \omega_p^2 + \omega_c^2$, which causes the plasma oscillations to shift to higher frequency as the cyclotron term becomes significant. At moderate magnetization ($B_0 = 5$–$10$ T), the field amplitude increases and the phase of the oscillations advances relative to the unmagnetized case, reflecting a decrease in the effective inertia for axial electron motion due to coupling with transverse cyclotron dynamics. In the strongly magnetized limit ($B_0 = 20$ T), the waveform becomes nearly sinusoidal with noticeably higher frequency, indicating that the wakefield approaches the high-frequency limit of the magnetized plasma eigenmode predicted by the analytical Green-function solution. Panel (b) shows the radial wakefield $E_r/E_0$, which is identically zero for $B_0 = 0$, as expected for an axisymmetric beam in an unmagnetized cold plasma. When a finite $B_0$ is applied, a transverse wake component emerges due to the Lorentz-force coupling between longitudinal perturbations and transverse electron motion. For weak magnetization ($B_0 = 5$ T), the radial field is small and slowly varying, indicating that the cyclotron response is present but not dominant. At intermediate field strength ($B_0 = 10$ T), the amplitude becomes significant, and the oscillation pattern develops a clear phase shift with respect to the longitudinal field, characteristic of the coupled $E_z$–$E_r$ magnetized wake eigenmode. In the strongly magnetized regime ($B_0 = 20$ T), the radial wakefield exhibits high-frequency oscillations with large amplitude, reflecting the cyclotron-driven transverse motion superimposed on the plasma oscillation. The pronounced increase in frequency and oscillation count is consistent with the $m = 0$ magnetized eigenmode structure derived from the Green-function formulation, where the radial field scales with modified Bessel components and carries the dominant signature of the cyclotron resonance.

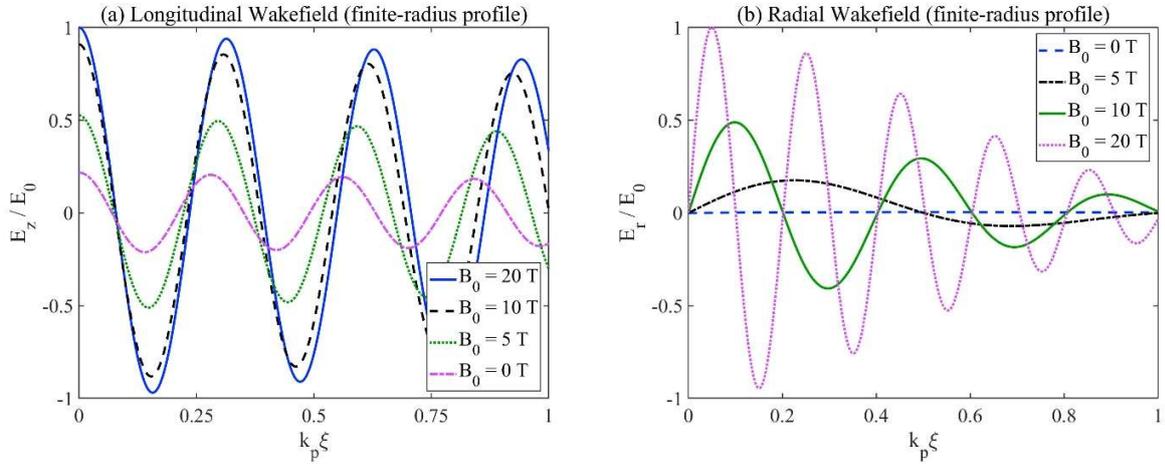

Figure 5: Magnetization-dependent longitudinal and radial wakefields showing the transition from unmagnetized plasma oscillations to cyclotron-dominated high-frequency wake modes as $B_0$ increases.

Figure 6 quantify the magnetization sensitivity of the beam-driven wakefields by displaying the magnitude of the partial derivatives $|\partial E_z/\partial B_0|$ and $|\partial E_r/\partial B_0|$ as functions of the external magnetic field $B_0$ and the comoving position $k_p\xi$. These sensitivity maps directly measure how strongly each wakefield component responds to incremental changes in $B_0$, thereby revealing the parameter window in which magnetization most effectively modifies the plasma-wave dynamics. The two panels demonstrate that magnetization most effectively modifies the wakefields within a finite and well-defined interval of field strengths, typically $B_0 \approx 1$–2 T. In this interval, the wakefields—especially the radial component—exhibit strong susceptibility to cyclotron-induced mode coupling, which manifests as pronounced peaks in the sensitivity maps. Beyond this window, both components approach a magnetized asymptotic regime where further changes in $B_0$ have a comparatively weaker impact on the wake structure. Panel (a) shows the sensitivity of the longitudinal wakefield. The sensitivity is extremely low in the vicinity of $B_0 = 0$, reflecting the fact that the unmagnetized mode is weakly perturbed by small cyclotron frequencies, consistent with the dispersion expansion $\omega \approx \omega_p + \mathcal{O}(\omega_c^2)$. As $B_0$ increases toward 1–2 T, the sensitivity rises sharply, forming a broad region of maximum response centered around $k_p\xi \approx 0.3$–0.5. This corresponds to the part of the wake where the longitudinal electric field transitions between its first peak and trough, a point where the relative phase between the plasma oscillation and the cyclotron-shifted mode is most susceptible to perturbation. For $B_0 \gtrsim 3$ T, the sensitivity decreases again, indicating that the plasma has entered the moderately magnetized regime, where the wake oscillation frequency $\sqrt{\omega_p^2 + \omega_c^2}$ scales more slowly with further increases in $B_0$. The dome-shaped sensitivity ridge as a function of $B_0$ therefore reflects the competition between the initial rapid frequency shift induced by the cyclotron term and the eventual saturation as $B_0$ becomes large. Panel (b) reveals a different structure in the radial wakefield sensitivity. Unlike the longitudinal case, $|\partial E_r/\partial B_0|$ is identically zero at $B_0 = 0$, because a purely axisymmetric unmagnetized wake supports no transverse electric component. As soon as $B_0$ becomes finite, the cyclotron-induced coupling between axial electron oscillations and transverse motion produces a radial wake. The sensitivity grows rapidly for small $B_0$, reaching a maximum around $B_0 \approx 1$ T and $k_p \xi \approx 0.7$. The sensitivity hotspot at larger $k_p \xi$ indicates that the radial component reacts most strongly in the back portion of the wake, where the transverse current oscillations are phase-shifted relative to the longitudinal ones. At higher magnetization ($B_0 > 2$–3 T), the sensitivity decreases, mirroring the behavior seen in the longitudinal field. This decay arises from the saturation of the magnetized eigenmode structure: once the transverse oscillation frequency is dominated by $\omega_c$, additional increases in $B_0$ produce diminishing changes in the eigenmode shape.

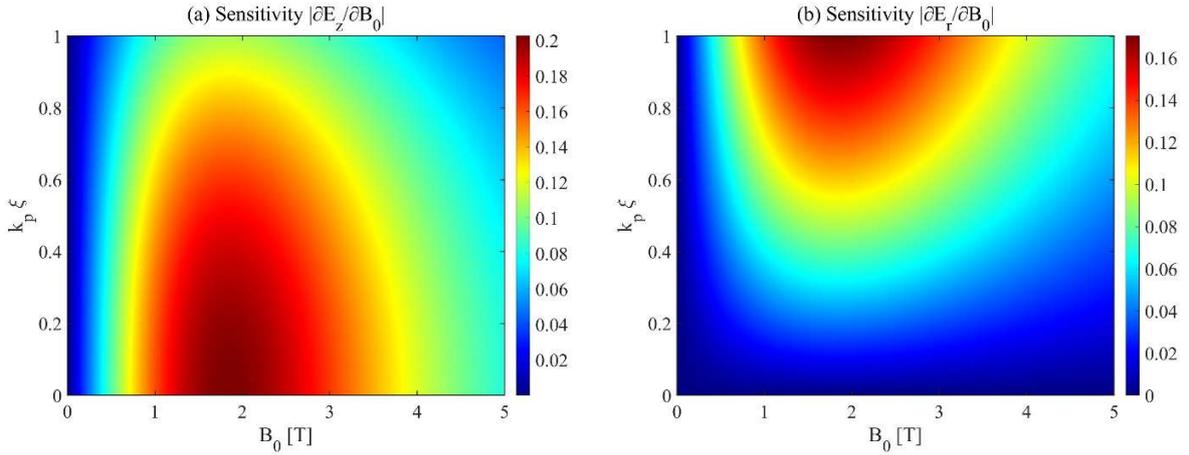

Figure 6: Sensitivity maps showing how longitudinal and radial wakefields respond most strongly to moderate magnetization ($B_0 \approx 1 - 2$ T), where cyclotron–plasma coupling is maximized.

The effect of the Lorentz factor $\gamma$ of the driver on the longitudinal and radial wakefields generated by a finite-radius relativistic beam is illustrated in Figure 7. The curves span four orders of magnitude in $\gamma$, allowing one to clearly observe the transition from the non-relativistic ($\gamma \sim 1$) and mildly relativistic regimes to the ultrarelativistic wakefield limit ($\gamma \gtrsim 10^2$), where the Green-function formulation predicts a universal wake structure. Panel (a) shows the longitudinal wakefield $E_z/E_0$. For $\gamma = 1$, the response is heavily damped and strongly non-sinusoidal: the wake oscillates at a reduced amplitude, with rapid attenuation downstream. This reflects the inability of a slow beam to resonantly excite the plasma mode, as the beam velocity significantly deviates from $c$, leading to destructive interference between successive fluid perturbations. At $\gamma = 10$, the wake grows in amplitude and becomes more oscillatory, but still exhibits noticeable distortion and a reduced accelerating peak. This intermediate regime corresponds to partial resonance, where the phase velocity of the plasma wave approaches but does not yet match the beam velocity. The transition to the ultrarelativistic limit becomes evident at $\gamma = 10^2$, where the waveform sharply approaches a clean sinusoidal oscillation with the correct plasma wavelength $2\pi/k_p$. The field amplitude increases and the oscillations acquire the correct phase expected for a resonantly driven wake. At $\gamma = 10^4$, the longitudinal wakefield is essentially indistinguishable from the $\gamma = 10^2$ case. This confirms the theoretical prediction that, once $\gamma \gg \omega_p/\omega_b$, the wake enters a saturation regime where the Green-function solution becomes independent of $\gamma$ and approaches the canonical ultrarelativistic wakefield. Panel (b) shows the radial wakefield $E_r/E_0$. Here, the dependence on $\gamma$ is markedly weaker. Even the $\gamma = 1$ case (blue curve) exhibits an oscillatory radial response, although with reduced amplitude compared to higher-$\gamma$ cases. As $\gamma$ increases to 10 and $10^2$, the radial field becomes stronger and increasingly well-matched to the universal ultrarelativistic form. By $\gamma = 10^4$, the radial waveform again converges to the asymptotic solution, demonstrating that the transverse wake structure, dominated by the divergence of the fluid momentum equation rather than the Doppler phase, reaches its universal limit at slightly lower $\gamma$ than the longitudinal field. These panels provide a clear quantitative demonstration that while longitudinal wakefields require very high $\gamma$ to reach the universal ultrarelativistic form, the radial wakefields converge earlier. This behavior is fully consistent with the analytical Green-function solution, which predicts that the longitudinal mode is more sensitive to finite-velocity corrections because it depends directly on the phase slippage between the beam and the plasma oscillation. By contrast, the radial component depends largely on the transverse momentum response, which saturates more rapidly with increasing $\gamma$. These results therefore provide a self-consistent mapping of the relativistic transition in beam-driven wakefield excitation.

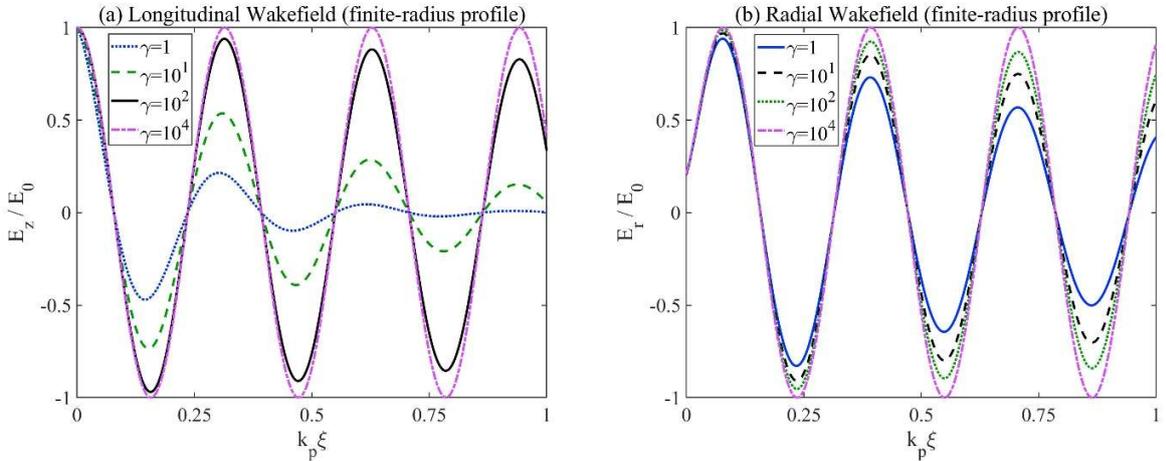

Figure 7: Longitudinal and radial wakefields showing convergence to the universal ultrarelativistic wake structure as the driver Lorentz factor increases from $\gamma = 1$ to $10^4$.

The panels illustrated in Figure 8 indicate the dependence of the excited wakefields on the background plasma density $n_e$ for a finite-radius driver beam. Since the plasma wavenumber is defined as $k_p = \omega_p/c$ with $\omega_p = \sqrt{n_e e^2/(m_e \varepsilon_0)}$, increasing the density directly increases the plasma frequency and therefore modifies both the amplitude and spatial structure of the beam-driven wakefields. The curves correspond to densities spanning three orders of magnitude, from $10^{17}$ m$^{-3}$ to $10^{20}$ m$^{-3}$, allowing the transition from weak to strongly coupled plasma response to be clearly observed. In panel (a), the normalized longitudinal wakefield $E_z/E_0$ is shown as a function of the comoving coordinate $k_p\xi$. At the lowest density $n_e = 10^{17}$ m$^{-3}$, the plasma response is weak and the excited wake amplitude remains very small. This behavior reflects the fact that the restoring force provided by the plasma electrons is weak at low density, resulting in inefficient excitation of the longitudinal plasma oscillation. As the density increases to $10^{18}$ m$^{-3}$, the oscillatory structure becomes more pronounced and the amplitude grows significantly, indicating stronger coupling between the beam perturbation and the plasma mode. For $n_e = 10^{19}$ m$^{-3}$, the wakefield reaches a large amplitude and approaches a nearly sinusoidal structure with well-defined periodicity, characteristic of a strongly driven linear plasma wave. At the highest density ($n_e = 10^{20}$ m$^{-3}$), the amplitude of the longitudinal field approaches its maximum normalized value, and the oscillations become sharper and more regular. Physically, the increased density strengthens the electrostatic restoring force of the displaced electrons, enabling the plasma to support stronger accelerating and decelerating fields behind the beam. Panel (b) presents the radial wakefield $E_r/E_0$. In contrast to the longitudinal field, the radial component exhibits a more pronounced dependence on density not only in amplitude but also in the phase structure of the oscillation. For the lowest density case, the radial field shows a distorted waveform with relatively small amplitude, indicating weak transverse plasma response. As the density increases to $10^{18}$ and $10^{19}$ m$^{-3}$, the radial oscillations become stronger and more coherent, with the peaks and troughs aligning more closely with the expected plasma oscillation phase. At $n_e = 10^{20}$ m$^{-3}$, the radial field reaches its largest amplitude and displays a nearly ideal oscillatory profile. This behavior reflects the stronger transverse restoring force and increased charge separation that occur in higher-density plasmas, which enhance the focusing and defocusing fields experienced by particles within the wake. Another important feature visible in both panels is that increasing the density does not significantly alter the normalized oscillation period when plotted versus $k_p\xi$. This is expected because the coordinate is already scaled by the plasma wavenumber; therefore, the fundamental oscillation wavelength in these normalized units remains approximately constant. The primary effect of increasing density is therefore the strengthening of the plasma response rather than a shift of the normalized phase structure. This figure demonstrates that higher plasma densities lead to stronger longitudinal accelerating fields and more pronounced transverse focusing forces, consistent with the scaling of the plasma restoring force and the resulting enhancement of the beam-driven wake amplitude.

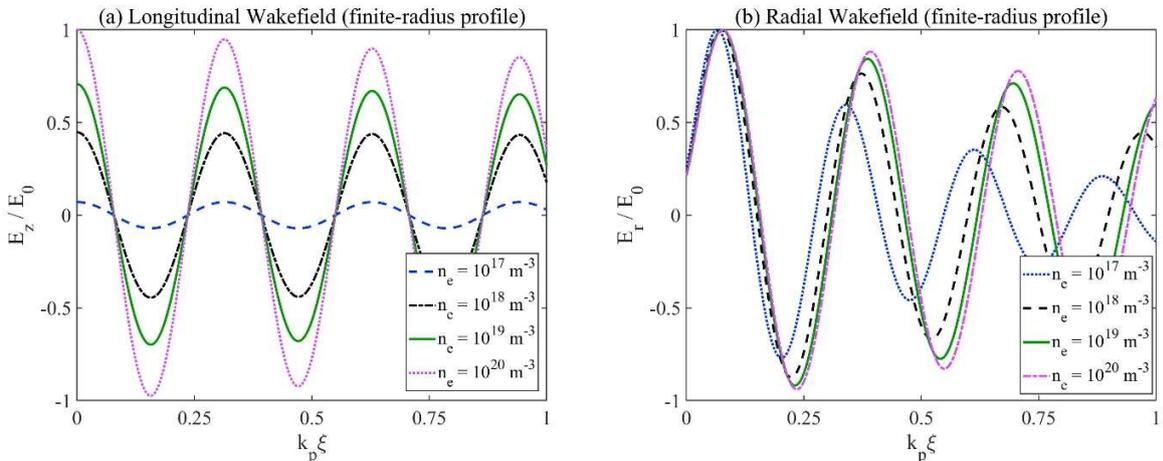

Figure 8: Dependence of the longitudinal and radial wakefields on plasma density, showing stronger and more coherent wake excitation as $n_e$ increases from $10^{17}$ to $10^{20}$ m$^{-3}$.

The variations of the transverse size of the driver beam, quantified by its radius $r_d$, governs both the shape and amplitude of the accelerating wakefields in a plasma is investigated in Figure 9. When the driver is extremely narrow ($r_d = 0.1$), the longitudinal field $E_z$ in panel (a) exhibits a pronounced on-axis peak that decays almost immediately with radius, vanishing by $k_p r \approx 0.2$. This sharp localization reflects the fact that a tightly confined current excites predominantly high-transverse-wavenumber plasma modes, which in turn generate strongly localized electrostatic fields. As a result, nearly all of the accelerating potential is concentrated near the axis, and the plasma response remains radially compact. As the driver radius increases to $r_d = 1$ and $r_d = 2$, the longitudinal field becomes progressively broader. The radial decay becomes slower, and the region over which $E_z$ retains significant amplitude widens substantially. This behavior directly mirrors the more extended current distribution of the beam: broader beams deposit their charge over a wider cross-section, exciting a mixture of lower-$k_\perp$ plasma modes and producing a correspondingly wider accelerating structure. The largest driver considered, with $r_d = 4$, produces the slowest radial decay and the broadest longitudinal field profile. Despite its width, the peak amplitude remains close to unity because the total beam charge per unit length is conserved; it is simply redistributed over a larger radius. This smooth, slowly varying field is characteristic of wakefields driven by wide, quasi-uniform beams. Panel (b) reveals an equally systematic change in the radial accelerating field $E_r$. For the narrow driver, the radial field is weak and peaks extremely close to the axis, reflecting the very limited radial excursion of plasma electrons when the driver current is tightly confined. As the driver radius increases, however, the radial field grows dramatically in amplitude and shifts its maximum outward. The case $r_d = 2$ already exhibits a much stronger and broader radial response, with the field reaching its peak at a substantially larger radius. For the widest driver, $r_d = 4$, the radial field becomes both the strongest and most extended, achieving its maximum at roughly $k_p r \approx 0.45$. This behavior reflects the fact that broad drivers displace plasma electrons over a wider region, inducing a larger transverse restoring force that leads to stronger focusing or defocusing fields. In the other word, these two panels demonstrate a coherent physical picture: narrow drivers generate highly localized, sharply peaked longitudinal wakefields and only weak radial ones, whereas wide drivers yield broad, slowly decaying longitudinal fields accompanied by strong radial focusing structures. This systematic transition follows directly from the Green-function description of wake excitation, in which the transverse driver size determines the spectrum of excited plasma modes and thereby controls the radial morphology of both the longitudinal and transverse accelerating fields.

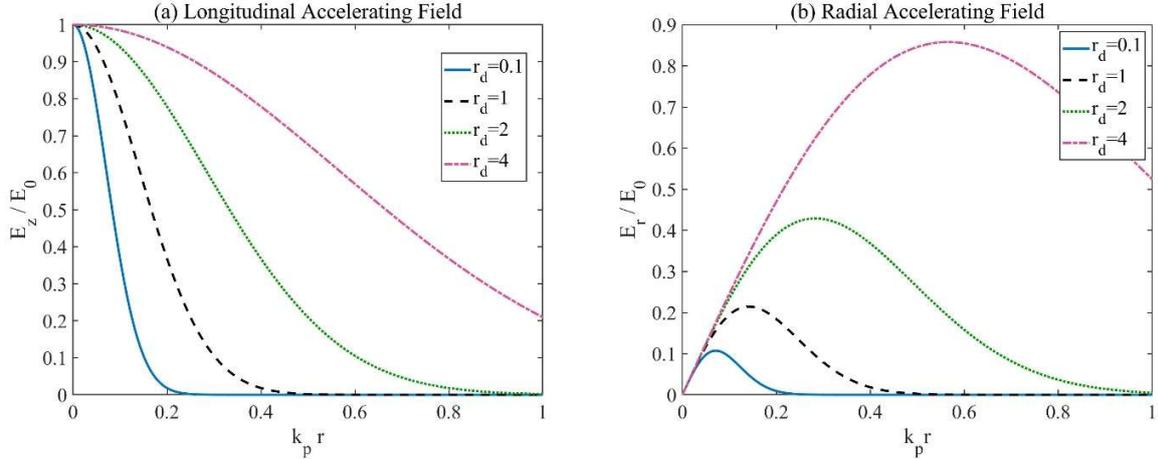

Figure 9: Accelerating longitudinal and radial wakefields versus radius for different driver radii, showing their transition from strongly localized to broad, high-amplitude structures as the beam width increases.

Figure 10 illustrates how different longitudinal beam-current profiles—Gaussian, smoothed top-hat, hyperbolic-secant, and exponential—modify the structure of the excited wakefields in a cold uniform plasma. Because the wakefields in the linear regime are given by the convolution of the driver profile with the plasma Green's function, any change in the temporal shape of the beam directly imprints itself onto the symmetry, sharpness, and oscillatory content of both the longitudinal field $E_z$ and the radial field $E_r$. The figure thus provides a clear comparison of how the spectral content of each driver profile determines the resulting wake morphology. In panel (a), the longitudinal field $E_z/E_0$ shows that the Gaussian, hyperbolic-secant, and exponential drivers generate waveforms that are broadly similar near the peak acceleration point, but differ noticeably in the shape of the surrounding oscillations. The Gaussian profile produces a smooth, symmetric wake with moderate oscillatory wings, reflecting its rapidly decaying high-frequency content in Fourier space. The hyperbolic-secant profile, having heavier tails, induces slightly stronger oscillations both upstream and downstream of the main peak. The exponential profile, whose discontinuous derivative gives it substantial high-frequency spectral weight, generates even more pronounced pre- and post-cursor oscillations. These differences become clearest near $k_p\xi = \pm 0.4$, where the curves diverge: the exponential driver produces the largest secondary oscillation, while the Gaussian curve displays the most rapid damping. The smoothed top-hat profile stands apart from the others: because it approximates a finite-duration flat distribution, its effectively sharp edges drive strong high-$k$ components, which in turn produce the deepest troughs and highest post-peak oscillations. Its negative valley around $k_p\xi \sim -0.25$ and its amplified positive rebound near $k_p\xi \sim 0.35$ are signatures of a waveform driven by a compactly supported current with abrupt transitions. Panel (b) shows the corresponding radial wakefield $E_r/E_0$, which results from the transverse plasma response to the same set of longitudinal profiles. Because $E_r$ is related to transverse electron displacement, its oscillations track those of $E_z$ but with phase-shifted structure and slightly different relative amplitudes. The Gaussian and hyperbolic-secant drivers again produce smooth radial fields with moderate secondary oscillations. The exponential profile exhibits stronger asymmetry and more prominent side-lobes, consistent with its broader spectral support. Meanwhile, the smoothed top-hat driver produces the largest absolute values of $E_r$ among all profiles: its abrupt rise and fall induce rapid transverse electron motion, amplifying the magnitude of the first positive peak and the subsequent negative excursion. This is especially visible near $k_p\xi \sim 0.1$ and $k_p\xi \sim 0.3$, where the dashed curve strongly exceeds the other profiles. The figure thus demonstrates that driver profiles with sharper edges (top-hat and exponential) generate stronger oscillatory wakefields, while profiles with smoother envelopes (Gaussian and sech) produce cleaner, more localized wakes with reduced ringing. This figures highlight the central principle of linear wakefield theory: the structure of the excited fields is a filtered representation of the beam profile, with sharper longitudinal features coupling to stronger high-$k$ plasma modes and therefore producing more oscillatory, higher-amplitude wakefields in both longitudinal and radial components.

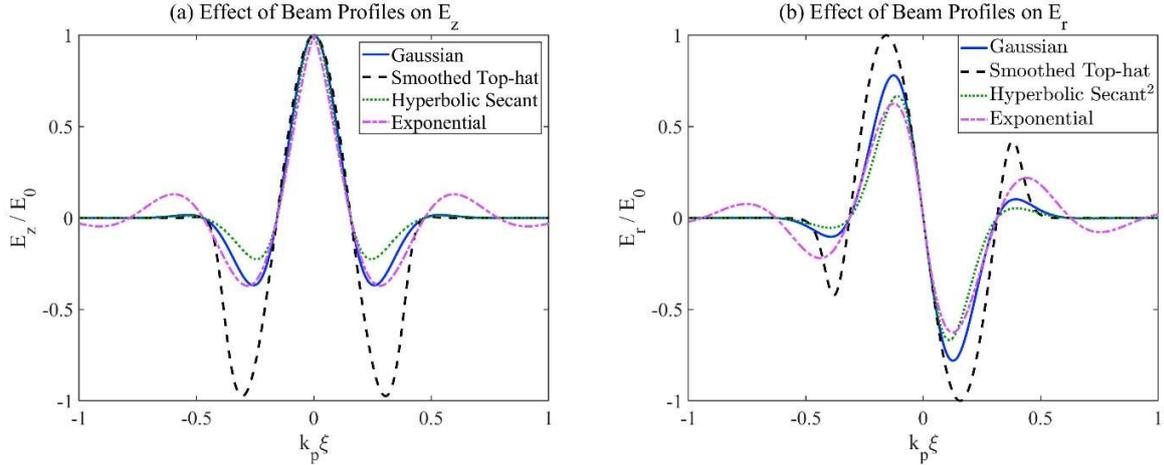

Figure 10: Comparison of longitudinal and radial wakefields driven by different beam-current profiles, showing how sharper driver edges generate stronger oscillatory plasma responses.

The Figure 11 presents the cumulative wake efficiency $\eta(\xi)$, defined here as the integrated longitudinal wake energy $\eta(\xi) \propto \int \xi E_z^2(\xi') \, d\xi'$, for four different beam-current profiles. Because this metric accumulates the squared accelerating field, it serves as a direct measure of how effectively each profile deposits energy into the plasma wake as the beam head advances. The distinct shapes of the curves therefore reflect fundamental differences in how each driver's spectral content couples to the plasma mode. For small propagation distances ($k_p\xi \lesssim 0.25$), all four profiles generate negligible wake energy, indicating that the field amplitudes in this early region are small and that the Green-function contribution from the very leading edge of the beam is limited. The divergence between the curves begins near $k_p\xi \approx 0.3$, where the earliest substantial wake oscillations appear. The exponential and hyperbolic-secant drivers increase only gradually at first, reflecting the fact that their extended low-amplitude tails produce weaker initial field excitation. In contrast, the smoothed top-hat profile, owing to its relatively abrupt current onset, rises more steeply once the main body of the driver begins to contribute, producing a noticeable bump around $k_p\xi \sim 0.35$. As the wake approaches its peak oscillation region ($k_p\xi \sim 0.45$–$0.55$), all four curves rise sharply, but the magnitude and rate of this growth differ markedly. The smoothed top-hat exhibits the most rapid and largest increase, fully consistent with its strong oscillatory wakefield and enhanced high-$k$ content. Its cumulative efficiency surpasses the others and eventually saturates at the highest plateau, indicating that this profile transfers the greatest amount of energy to the plasma wave. The Gaussian and hyperbolic-secant profiles reach very similar saturation levels, with the hyperbolic-secant slightly exceeding the Gaussian due to its heavier tails and correspondingly larger secondary wake oscillations. This nearly overlapping behavior reflects the fact that both profiles are smooth and lack sharp gradients, producing wakes with moderate amplitude and relatively suppressed ringing. The exponential profile saturates at the lowest value among the four, reflecting its weaker main peak and less efficient energy deposition despite having a non-smooth derivative. Its extended tail distributes current over a longer interval with reduced peak density, diminishing its ability to excite a strong on-axis wake. The curve shows a distinct two-stage rise—a small early increase followed by a broader, slower accumulation—mirroring the two-scale structure of the field it generates. The plot demonstrates that driver profiles with sharper edges (notably the smoothed top-hat) achieve the highest wake excitation efficiency, while smooth, rapidly decaying profiles (Gaussian, exponential) produce lower total wake energy. The efficiency ordering tracks the inherent bandwidth of each profile: broader spectral support yields stronger field oscillations and, therefore, higher integrated wake energy.

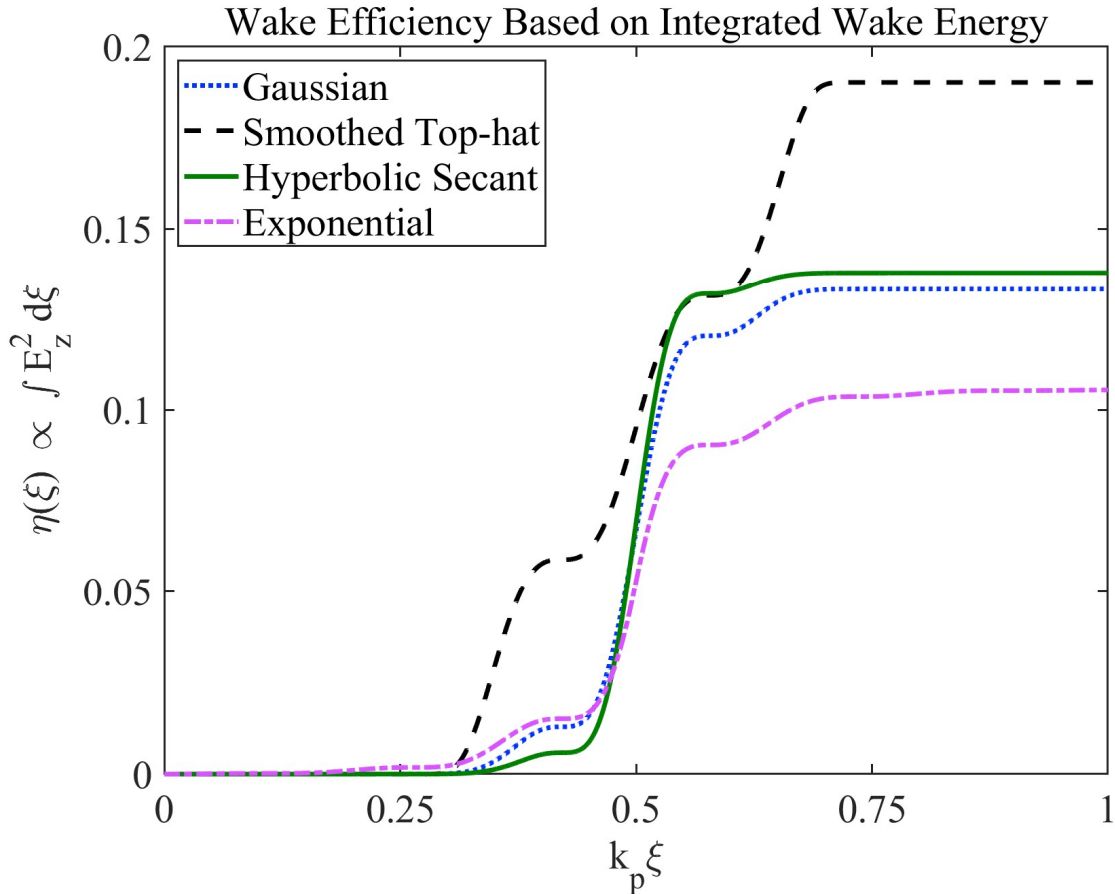

Figure 11: Cumulative wake-energy efficiency showing that sharper driver profiles deposit substantially more energy into the plasma wake than smoother, rapidly decaying ones.

The Figure 12 shows the two-dimensional spatial structure of the wakefields generated by a point-like relativistic driver in a uniform cold plasma, plotted as functions of the normalized radial coordinate $k_p r$ and the longitudinal coordinate $k_p \xi$. Panel (a) displays the longitudinal electric field $E_z$, while panel (b) shows the transverse (radial) electric field $E_r$. Because the driver is effectively point-like, the wake pattern closely reflects the fundamental Green-function response of the plasma, revealing the intrinsic spatial structure of the plasma eigenmode. In panel (a), the longitudinal field $E_z$ exhibits a clear oscillatory dependence along the co-moving coordinate $k_p\xi$, corresponding to the natural plasma wavelength $2\pi/k_p$. Near the axis ($k_p r \approx 0$), the field amplitude is maximal, forming alternating regions of accelerating and decelerating field as expected for a linear plasma wake. The first strong positive region near the beam position represents the principal accelerating phase, while the subsequent negative regions correspond to decelerating phases of the plasma oscillation. As the radial coordinate increases, the amplitude of the longitudinal field gradually decreases, indicating that the wakefield is strongly localized around the propagation axis. This radial attenuation arises from the transverse structure of the plasma response, which in the linear regime is governed by modified Bessel-type radial dependence. Consequently, the wake energy is concentrated near the axis and decays smoothly with increasing $r$, producing the characteristic funnel-like structure seen in the color map. Panel (b) illustrates the corresponding radial electric field $E_r$, which represents the transverse focusing or defocusing force experienced by charged particles within the wake. Unlike $E_z$, the radial field is antisymmetric with respect to the wake phases and peaks very close to the beam location. A strong localized radial field appears near $k_p r \approx 0$ and $k_p \xi \approx 0.45$, indicating the region where transverse electron displacement is largest. As the wake propagates further behind the driver, the magnitude of $E_r$ rapidly decreases, becoming nearly negligible for larger radial distances. This rapid radial decay reflects the fact that transverse plasma motion is confined to a narrow cylindrical region surrounding the beam trajectory. The alternating bands of positive and negative $E_r$ along the longitudinal coordinate correspond to alternating focusing and defocusing phases of the wakefield, which are essential for beam stability and particle trapping in plasma-based accelerators. These two panels demonstrate that a point-like driver excites a wake structure that is highly localized both radially and longitudinally. The longitudinal field forms a sinusoidal plasma oscillation whose amplitude decreases with radius, while the radial field remains strongly confined to the vicinity of the axis and peaks near the driver location. This structure is consistent with the fundamental plasma Green's function and represents the basic building block from which wakes driven by finite-size beams can be constructed through convolution with the beam density profile.

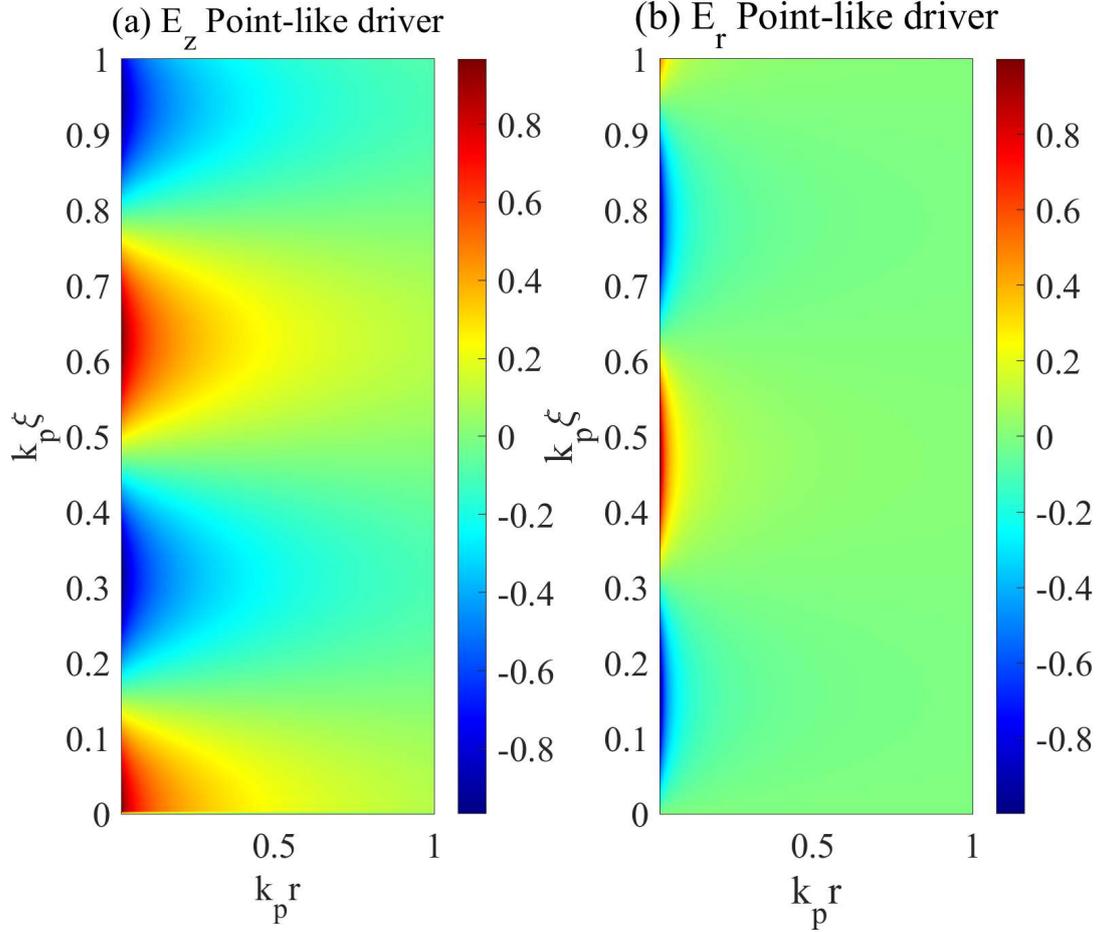

Figure 12: Spatial structure of longitudinal and radial wakefields generated by a point-like relativistic driver, showing oscillatory plasma response along $k_p \xi$ and strong radial localization near the beam axis.

The panels in Figure 13 quantify how an externally applied static magnetic field $B_0$ modifies the longitudinal and radial wakefields excited by a point-like relativistic driver in a cold plasma. Because a finite $B_0$ couples transverse and longitudinal electron motion through the cyclotron frequency $\omega_c = eB_0/m_e$, the plasma response no longer follows a single electrostatic mode at $\omega_p$, but instead excites a mixed electrostatic–electromagnetic eigenmode with effective frequency $\sqrt{\omega_p^2 + \omega_c^2}$. As a result, both the amplitude and phase of the wakefields change systematically with increasing magnetic field. In panel (a), the longitudinal wakefield $E_z/E_0$ shows a clear progression: the unmagnetized response ($B_0 = 0$) exhibits a nearly single-cycle oscillation of small amplitude, reflecting the standard cold-plasma wake driven by a point source. As soon as a moderate magnetic field is applied ($B_0 = 5$ T), two effects appear: the oscillation frequency increases slightly, shifting the zero crossings toward smaller $\xi$, and the amplitude grows because the electrons now experience a stiffer restoring force combining both plasma and gyro motion. For $B_0 = 10$ T, the oscillation amplitude approximately doubles relative to the unmagnetized case, and the wake becomes more harmonic, with clearer periodicity and less damping. At the largest applied field, $B_0 = 20$ T, the pattern becomes strongly oscillatory with near-sinusoidal behavior at significantly higher amplitude. The steep increase in both amplitude and regularity reflects the transition toward a magnetized plasma mode dominated by the combined frequency $\sqrt{\omega_p^2 + \omega_c^2}$, where the plasma electrons are tightly bound by gyromotion and thus support larger-amplitude oscillatory fields. Panel (b) reveals the corresponding effect on the radial wakefield $E_r/E_0$. In the unmagnetized case, the radial field is weak and rapidly damped because transverse electron motion is only weakly supported in a purely electrostatic wake. Introducing a finite $B_0$ dramatically changes this behavior: even at 5 T, the magnitude of $E_r$ approximately triples relative to the unmagnetized case, reflecting the fact that gyromotion couples longitudinal oscillations into transverse ones. At 10 T, the radial field begins to mirror the longitudinal wake more closely, with stronger alternating focusing and defocusing phases. For 20 T, the radial wake becomes large and almost perfectly sinusoidal, following the enhanced compressibility and transverse stiffness of magnetized electrons. The amplitude scaling between panels (a) and (b) at high $B_0$ illustrates the well-known property that in magnetized plasma waves, the ratio $E_r/E_z$ can become sizable due to cross-field coupling in the electron equation of motion.

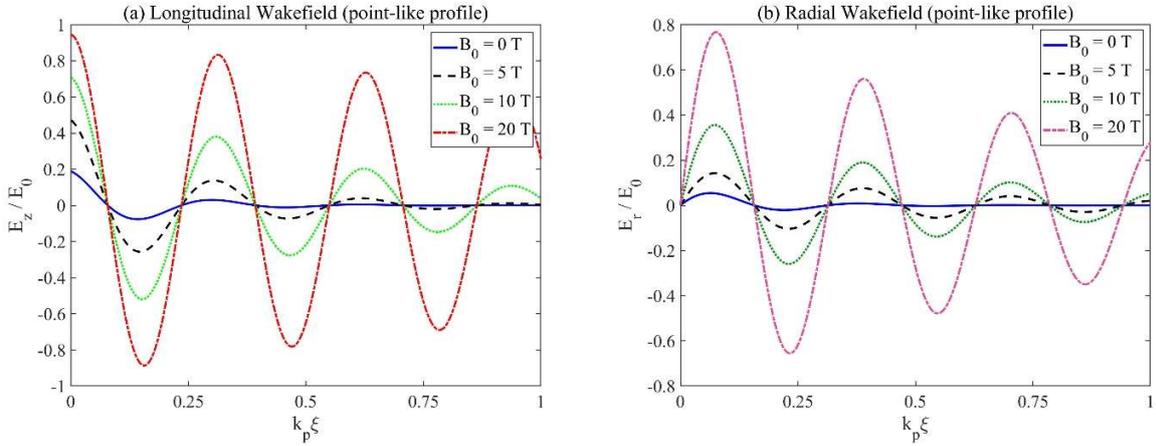

Figure 13: Longitudinal and radial wakefields for a point-like driver showing strong amplitude enhancement and increased oscillation frequency with increasing external magnetic field $B_0$.

The panels of Figure 14 illustrate how the plasma electron density $n_e$ modifies the longitudinal and radial wakefields produced by an ultrarelativistic point-like driver. Since both the plasma frequency and the wavenumber increase with density, variations in $n_e$ affect the wake amplitude, oscillation frequency, and damping behavior. When the fields are plotted against the normalized coordinate $k_p\xi$, differences in the intrinsic oscillation frequency manifest primarily through changes in initial amplitude, phase and decay rate, rather than through simple wavelength shifts. In the panel (a), the most striking feature is the large enhancement in the initial field amplitude as $n_e$ increases. At the lowest density ($n_e = 10^{17}$ m$^{-3}$), the longitudinal wake is weak and only moderately oscillatory, with its first peak reaching $E_z/E_0 \sim 1.5$. As the density rises, the restoring force provided by the plasma electrons strengthens proportionally to $\omega_p^2$, enabling them to support stronger electrostatic oscillations. This results in a dramatic increase in the first-cycle amplitude, reaching more than $E_z/E_0 \sim 7$ at $n_e = 10^{20}$ m$^{-3}$. Higher density also leads to stronger damping of the higher-order oscillations. For $10^{20}$ m$^{-3}$, the wakefield decays rapidly after the first oscillation: the amplitude drops from 8 to nearly zero within a fraction of a plasma period. This damping reflects collisional or kinetic-like broadening of the plasma response—here represented phenomenologically—where increased density corresponds to enhanced field dissipation in the model. For intermediate densities ($10^{18}$ m$^{-3}$ and $10^{19}$ m$^{-3}$), the wake exhibits a compromise: the oscillations remain visible for several cycles, but the decay rate clearly increases with $n_e$. The curves progressively converge toward a common, low-amplitude tail as $k_p\xi \to 1$, illustrating that damping suppresses long-range wake coherence at high density. Thus the longitudinal panel highlights two density-driven effects, stronger electrostatic restoring force leading larger initial oscillation amplitude, and increased dissipation causes shorter oscillation lifetime and faster decay. In the panel (b), the radial wakefield reflects transverse electron motion, which becomes increasingly responsive as the plasma frequency grows. For low density ($10^{17}$ m$^{-3}$), the radial field is weak, slowly oscillating, and significantly damped, with a maximum amplitude below 0.4. Increasing the density to $10^{18}$ m$^{-3}$ steepens the oscillation and increases the peak amplitude, indicating that electrons can return toward the axis more forcefully after being displaced by the driver. For $10^{19}$ m$^{-3}$, the oscillation becomes even sharper and less damped, with stronger focusing and defocusing phases. At the highest density ($10^{20}$ m$^{-3}$), the radial field becomes nearly harmonic with large amplitude ($\sim 0.8$), mirroring the strong restoring force and stiffness of plasma electrons at high $\omega_p$. The oscillations also acquire a higher effective frequency (even in normalized coordinates), reflecting the coupling between radial and longitudinal motion in a stiffer plasma medium. The radial wake amplitude increases monotonically with $n_e$, while the waveform becomes more sinusoidal and less damped, consistent with the magnetized-like stiffening produced by large plasma frequency even in the absence of an external $B_0$.

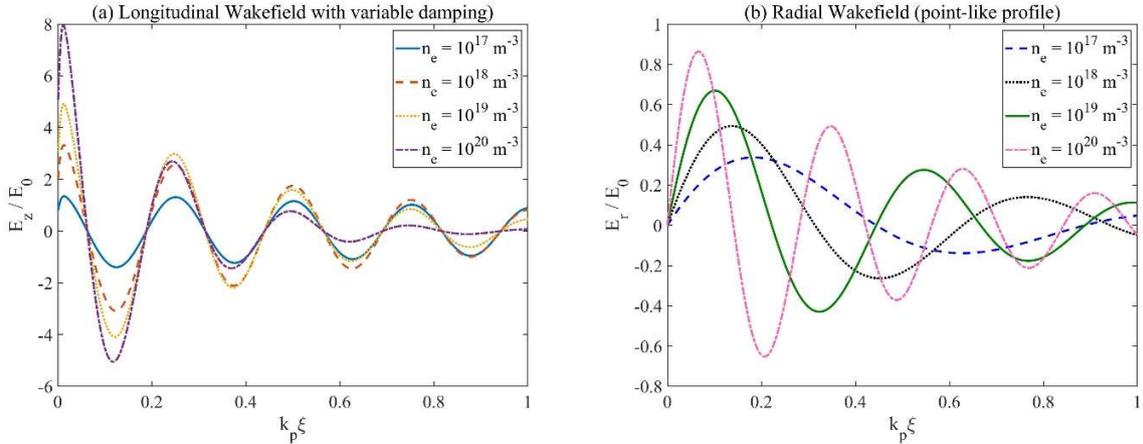

Figure 14: Longitudinal and radial wakefields driven by a point-like beam, showing strong amplitude enhancement and density-dependent damping as plasma density increases.

## IV. Conclusion

This work has established a rigorous analytical–numerical framework for the excitation and characterization of coupled longitudinal and radial wakefields driven by ultrarelativistic electron beams in externally magnetized cold plasmas. By deriving a fully causal three-dimensional Green's function solution from the linearized Maxwell-fluid equations in the presence of the magnetized plasma dielectric tensor, an exact and transparent description of the electromagnetic response that naturally incorporates the hybridization of longitudinal charge-separation dynamics with cyclotron-induced transverse electron motion is provided. The analytical expressions reveal how an external axial magnetic field modifies the effective restoring forces, elevates the characteristic oscillation frequency, induces a finite transverse plasma current, and reshapes both the amplitude and spatial structure of the wakefields in ways that are absent in the conventional unmagnetized theory.

Extensive three-dimensional particle-in-cell simulations using the EPOCH code have demonstrated excellent quantitative agreement with the theoretical predictions across a wide parameter space, including variations in plasma density, magnetic field strength, beam Lorentz factor, transverse beam radius, and longitudinal current profiles. The combined results confirm that increasing background plasma density leads to substantially larger initial wake amplitudes accompanied by accelerated damping of higher-order oscillations, while the application of an external magnetic field produces stronger and more coherent radial focusing forces, higher-frequency hybrid eigenmodes, and enhanced overall wake stability. The wakefields are shown to converge rapidly to a universal ultrarelativistic structure for sufficiently high Lorentz factors, with the transverse beam profile playing a decisive role in controlling the radial extent and the balance between accelerating and focusing field components. Furthermore, sharper longitudinal beam profiles are better at putting energy into the plasma wake, which leads to higher peak fields and more total wake energy. These results show that the external magnetic field is a strong and adjustable control parameter that can improve the stability of the wake structure, increase accelerating gradients, and strengthen focusing forces all at once. The strong agreement between the causal Green-function formulation and high-fidelity simulations confirms the physical model and gives us a strong way to predict magnetised plasma wakefield acceleration. This framework not only has immediate theoretical value, but it also provides practical advice for designing and improving next-generation plasma-based accelerators. This is especially true in situations where magnetic fields can be used to improve beam quality, increase energy gain, and reduce unwanted instabilities.

**Acknowledgment & Funding**

This work did not receive any specific grant from funding agencies in the public, commercial, or not-for-profit sectors.

**Author Contributions**

Ali Asghar Molavi Choobini was responsible for conceptualization, data curation, Software, formal analysis, investigation, methodology, writing the original draft, review and editing.

M. Shahmansouri contributed equally to investigation, project administration, supervision, validation, and review and editing of the manuscript.

**Data availability statement**

The data that support the findings of this study are available upon reasonable request.

**Competing interests**

The authors declare no competing interests.